\documentclass{elsarticle}
\usepackage[a4paper, left = 2.1cm, right = 2.1cm, top = 2.25cm, bottom = 2.05cm]{geometry}
\usepackage{amsmath}
\usepackage{amssymb}

\usepackage{slashed}

\usepackage{graphicx}
\usepackage{subfigure}
\usepackage{color}

\begin{document}

\begin{frontmatter}

\title{{\bf Majorana Phase And Matter Effects In Neutrino Chiral Oscillation}}

\author[1,2]{Ming-Wei Li}
\ead{limw2021@sjtu.edu.cn}
\author[1,2]{Zhong-Lv Huang\corref{cor1}}
\ead{huangzhonglv@sjtu.edu.cn}
\author[1,2,3]{Xiao-Gang He}
\ead{hexg@sjtu.edu.cn}

\cortext[cor1]{Corresponding author}

\address[1]{{Tsung-Dao Lee Institute, and School of Physics and Astronomy, Shanghai Jiao Tong University},  {Shanghai}, {200240}, {China}}

\address[2]{{KLPAC and SKLPPC Laboratories, Shanghai Jiao Tong University}, {Shanghai}, {200240}, {China}}

\address[3]{{Department of Physics, National Taiwan University}, {Taipei}, {10617}, {Taiwan}}

\begin{abstract}
Due to finite masses and mixing, for neutrinos propagation in space-time, there is a chiral oscillation between left- and right- chiral neutrinos,  besides the usual oscillation between different generations. The probability of chiral oscillation is suppressed by a factor of $m^2/E^2$ making the effect small for relativistic neutrinos. However, for non-relativistic neutrinos, this effect can be significant. In matter, the equation of motion is modified. When neutrinos produced in weak interaction pass through the matter, the  eigen-energies are split into two different ones depending on the helicity of the neutrino. This results in different oscillation behavior for neutrinos with different helicity, in particular there is a new resonant effect related to the helicity state of neutrino different than the usual MSW effect. For Majorana neutrinos, chiral oscillation also depends on Majorana phases.
\end{abstract}

\end{frontmatter}

\section{Neutrino chiral oscillation}
\label{introduction}
Neutrino oscillation has been observed experimentally \cite{ParticleDataGroup:2022pth}. Most of the studies have concentrated on neutrino oscillations between generations with the same chirality. In this paper we report our new and interesting results on neutrino oscillations between different chiral states. In the standard model (SM), active neutrinos have left-chiral interaction and therefore neutrinos produced are left-chiral states. Due to finite masses, the left-chiral neutrinos could oscillate into right-chiral ones as pointed out in Refs \cite{Bittencourt:2020xen,Ge:2020aen,Kimura:2021qlh,Bittencourt:2022hwn,Blasone:2022fgj}. Extend the discussion for neutrinos propagate in matter, there are further modifications for neutrino oscillation behavior.
In matter, each of the  eigen-energy is split into two different ones depending on the helicity of the neutrino \cite{Pantaleone:1992xh}. We find that matter effects result in different oscillation behavior for neutrinos with different helicity, in particular there is a new resonant effect related to the helicity of neutrino different than the MSW effect \cite{Wolfenstein:1977ue,Mikheyev:1985zog}. For Majorana neutrinos, neutrino chiral oscillation in matter can reveal information about Majorana phases which is drastically different from the usual oscillation. These effects provide a new insightful understanding of neutrino oscillation and neutrino properties, and have profound impacts on cosmic and astro neutrinos physics. We provide some details in the following.

The evolution of neutrinos in free space propagation is governed by the Dirac equation
\begin{equation}
	(i \slashed{\partial}- m) \psi = 0\;,
\end{equation}
which gives a Hamiltonian $
%H=\gamma^0\boldsymbol{\gamma}\cdot\mathbf{p}+m\gamma^0
H =(\mathbf{p}\cdot\boldsymbol{\Sigma})\gamma^5+m\beta$. Here $\boldsymbol{\Sigma}= \gamma^5\gamma^0\boldsymbol{\gamma} $ is the spin  operator and $\beta=\gamma^0$.

In the Schr\"odinger picture, the Hamiltonian operator generates the time evolution operator of quantum states,
\begin{eqnarray}
	U(t)=e^{-iHt} 
	=\cos(Et)-i{(\mathbf{p}\cdot\boldsymbol{\Sigma})\gamma^5+m\beta\over E} \sin(Et)\;,
\end{eqnarray}
and the wave function evolves as $\psi(t) = U(t)\psi(0)$.  We have used the chiral representation for the $\gamma^\mu$ matrices,
\begin{equation}
	\gamma^0=\begin{pmatrix}
		0&I\\
		I&0
	\end{pmatrix}\;,\;\;
	\gamma^i=\begin{pmatrix}
		0&\sigma^i\\
		-\sigma^i&0
	\end{pmatrix}\;,\;\;
	\gamma^5=\begin{pmatrix}
		-I&0\\
		0&I
	\end{pmatrix}\;.\;\;
\end{equation}

The positive energy and negative energy $\bar E$ wave functions are given by
\begin{equation}
	\begin{aligned}
		\psi^{h(\bar E=+E)}(\mathbf{x}) =
		\frac{1}{\sqrt{2E}}
		\begin{pmatrix}
			\sqrt{E-h\cdot p}\ u^h\\
			\sqrt{E+h\cdot p}\ u^h
		\end{pmatrix}e^{i\mathbf{p}\cdot\mathbf{x}}\;,\;\;
		\psi^{h(\bar E= -E)}(\mathbf{x}) = 
		\frac{1}{\sqrt{2E}}
		\begin{pmatrix}
			\sqrt{E+h\cdot p}\ u^h\\
			- \sqrt{E-h\cdot p}\ u^h
		\end{pmatrix}e^{i\mathbf{p}\cdot\mathbf{x}}\;,
	\end{aligned}
\end{equation}
where $E=\sqrt{p^2 +m^2}$, 
$ h=\pm1 $ and $u^h$ are the helicity states with  $(\mathbf{p}\cdot\boldsymbol{\Sigma})u^h = (h\cdot p) u^h$. The wave function is normalized as $\int\mathrm{d}x\left( \psi'(\mathbf{x})\right)^\dagger \psi(\mathbf{x})/V =\delta_{p'p}$.

The wave function $\psi^{h}(t,\mathbf{x})$, evolved from the momentum eigenstate  $\psi^h(\mathbf{x})$ produced at $t=0$ with momentum $\mathbf{p}$, will be $
\psi^{h}(t,\mathbf{x})=U(t)\psi^h(\mathbf{x})= \psi^h(\mathbf{x}) e^{-i\bar Et}$. This is an more effective way of obtaining wave function time evolution compared with the methods used in previous studies \cite{Bittencourt:2020xen,Ge:2020aen,Kimura:2021qlh,Bittencourt:2022hwn,Blasone:2022fgj}.

The wave function $\psi$ can be decomposed into left-chiral $\psi_L = L\psi$ and right-chiral $\psi_R= R \psi$ with $L = (1-\gamma_5)/2$ and $R=(1+\gamma_5)/2$, so that $\psi = \psi_L + \psi_R$. The left-chiral and right-chiral neutrinos are entangled during propagation by
$ i \slashed{\partial}\psi_L = m \psi_R$ and $ i \slashed{\partial}\psi_R = m \psi_L$. 
If a pure $\psi_L$ is produced, like the standard weak interaction neutrino production, at some later time, $\psi_L$ will oscillate into $\psi_R$ or vice versa.  At $ t=0 $, a left-chiral normalized neutrino wave function would be
\begin{equation}
	\psi^h_{L}(\mathbf{x})
	=\begin{pmatrix}u^h\\0\end{pmatrix}
	e^{i\mathbf{p}\cdot\mathbf{x}}\;,\;\;
	\psi^h_{R}(\mathbf{x})
	=\begin{pmatrix}0\\u^h\end{pmatrix}
	e^{i\mathbf{p}\cdot\mathbf{x}}\;.
\end{equation}
At time $ t $, 
$\psi^h_{L}(t,\mathbf{x})$  would be
$	\psi^h_{L}(t,\mathbf{x})=U(t)\psi^h_{L}(\mathbf{x})
=\left( \cos(Et)+i(h\cdot p/E) \sin(Et)\right)\psi^h_{L}(\mathbf{x})
-i (m/E) \sin(Et)\psi^h_{R}(\mathbf{x})
$.

When probed at time $ t $ by the normalized the  state $\psi^h_{L(R)}$ at $ \mathbf{x}=\mathbf{L} $, i.e. $\psi^h_{L}(\mathbf{x-l})=\psi^h_{L}(\mathbf{x})e^{-i\mathbf{p}\cdot\mathbf{L}}$ 
or $\psi^h_{R}(\mathbf{x})e^{-i\mathbf{p}\cdot\mathbf{L}}$. However, under the same momentum assumption one can ignore the phase $e^{-i\mathbf{p}\cdot\mathbf{L}}$. The chiral neutrino oscillation probabilities would be
\addtocounter{equation}{1}
\begin{align}
	\label{free}
	&P(\nu^h_L\rightarrow\nu^h_L)=|\left\langle\psi_{L}^h(\mathbf{x})|\psi^h_{L}(t,\mathbf{x})\right\rangle|^2= 1-{m^2\over E^2} \sin^2(Et)\;,\tag{\theequation.1}\\
	\label{freee}
	&P(\nu^h_L\rightarrow\nu^h_R)=|\left\langle\psi_{R}^h(\mathbf{x})|\psi^h_{L}(t,\mathbf{x})\right\rangle|^2= {m^2\over E^2} \sin^2(Et)\;.\tag{\theequation.2}
\end{align}
One sees that a pure left-chiral state has partially evolved into a right-chiral state. A chiral oscillation has occurred.

For high energy neutrinos with $m/E\ll1$, the probability for chiral oscillation is small, and the oscillation is the same as the Dirac neutrino $\psi^h$ justifying the usual neutrino oscillation treatment. When generation mixing is included, one obtains the usual neutrino oscillation probability formula. But for non-relativistic neutrinos, like cosmic relic neutrinos whose energies are very low, the chiral oscillation probability can reach 1/2 on the average. The chiral oscillation has a significant consequence on the detection of the cosmic neutrino background even for one generation Dirac neutrino~\cite{Ge:2020aen}.

In reality, neutrinos are produced through weak interaction which is left-chiral and the resulting neutrinos are dominated by left-chiral ones. Let us take the $\nu_\mu$ produced in $\pi^+ \to \mu^+ \nu_\mu$ as an example to demonstrate this. The effective weak interaction for this process is $~ (\bar d \gamma^\mu L u) (\bar \nu \gamma_\mu L \mu)$. Taking the matrix element $\langle0|\bar d \gamma^\mu L u |\pi^+\rangle\sim i f_\pi p^\mu_\pi$, one obtains $M(\pi^+ \to  \mu^+   \nu_\mu)
\sim i f_\pi \bar \nu (m_{\nu} L - m_\mu R) \mu$. We see that $-i f_\pi m_\mu \bar \nu R \mu$ operator term makes a leading contribution, which generates left-chiral neutrino.
So practically the neutrinos generated in this decay is a purely left-chiral. Therefore, Eq.(\ref{free}) and Eq.(\ref{freee}) describe in practice what happen to neutrino chiral oscillation.

Note that detection at $\mathbf{x}=\mathbf{L} $ is also through weak interaction, it is therefore sensitive to detect left-chiral neutrinos and the right-chiral neutrinos will practically be not detectable. Therefore, to experimentally known chiral oscillation occurred will be through Eq.(\ref{free}) which represents left-chiral neutrino disappearance probability,  not by Eq.(\ref{freee}) which represents appearance of right-chiral neutrino. In our later discussions we will work out disappearance and appearance probabilities to keep track of what happen to different chirality of neutrinos, but with the understanding that only left-chiral neutrino disappearance probability can be probed by SM weak interactions. Particularly, when neutrinos are Majorana ones, the left-chiral component and right-component isn't independent. That is to say, the charge conjugate of left-chiral neutrinos is right-chiral ones. So the left-chiral to right-chiral neutrino appearance processes can also be measured.

We also would like to mention that the same chiral oscillation can happen between left- and right- chiral charged lepton. In these cases the oscillation length is very short, for example the lightest charged lepton, electron, one period is $t_p = 2\pi/E_e < 2\pi/m_e =8.1\times 10^{-21} \mathrm{s}$. Such a short period oscillation cannot be observed. The chiral oscillation is always time average one with a probability of $m^2_e/2E_e^2$. For neutrinos, because the rest masses are much smaller, oscillation length for relic cosmic neutrinos, the period $t_p = 2\pi/E_\nu$ can be at the ps range. We hope in the future chiral oscillation can be detected with new ultra-fast detection techniques. We however, would like to point out that even with time average detections, for Majorana neutrinos, there are still observation effects if there are non-zero Majorana phases.
	
\section{The matter effects}
For neutrinos propagating in matter, due to $W$ and $Z$ exchanges, neutrinos interaction with electron, proton and neutron in SM will change the behavior of neutrino oscillation. The effective interaction Lagrangian in matter is given by
$	\mathcal{L}_{\mathrm{int}} = 
-\bar{\nu}_L  j_L^\mu\gamma_\mu  \nu_{L}\;. 
$
Here $j_L^\mu$ is the effective matter current which neutrino can interact.  When studying neutrino oscillation in matter, one should include $\mathcal{L}_{\mathrm{int}}$. In the rest frame of the homogeneous, isotropic, unpolarized, electrical neutrality medium, $j_L^\mu = (\rho, \mathbf{0})$, and the $Z$ contribution from electron and proton would cancel out so that $ \rho$ is a diagonal matrix $\operatorname{diag}\left\{\rho_W+\rho_Z, \rho_Z, \rho_Z \right\} $ with $ \rho_W=\sqrt{2}G_F N_e $ and $ \rho_Z=-G_F N_n/\sqrt{2} $ where $N_e, N_n$ are the number density of electron and neutron so that the elements of $\rho$ could be positive or negative. In our later discussions, we will work in the frame where the matter is at rest.

Neutrinos may be Majorana particles such as those in seesaw models. 
The most general effective Lagrangian for these seesaw neutrinos in matter will be given by 
	\begin{eqnarray}
		\label{general lagrangian}
		\begin{aligned}
			\mathcal{L}=&\bar\nu_{L} i\slashed{\partial} \nu_{L} 
			+ \bar{N}_{R} i\slashed{\partial} N_R
			-\frac{1}{2}\left( \begin{pmatrix}
				\bar{\nu}^c_L & \bar{N}_R
			\end{pmatrix}
			\begin{pmatrix}
				M_L & M_D^T\\ M_D & M_R
			\end{pmatrix}
			\begin{pmatrix}
				\nu_L\\ N_R^c
			\end{pmatrix}+\text{h.c.}\right)
			-\begin{pmatrix}
				\bar{\nu}_L & \bar{N}_R^c
			\end{pmatrix}
			\begin{pmatrix}
				j_L^\mu & j_{RL}^\mu\\  j_{RL}^{\mu\dagger} & j_R^{\mu}
			\end{pmatrix}\gamma_\mu
			\begin{pmatrix}
				\nu_L\\ N_R^c
			\end{pmatrix}\\
			=&\bar \psi_L i\slashed{\partial} \psi_L
			-\frac{1}{2}\left(\bar\psi_L^c \mathcal{M}\psi_L+\text{h.c.}\right)
			-\bar{\psi}_{L}J^\mu\gamma_\mu\psi_L\;.
		\end{aligned}
	\end{eqnarray}
$j^\mu_{RL}$ and $j^\mu_R$ are beyond the SM contributions which vanish in Type I, II and III seesaw models. $N_R$ are possible right-chiral neutrinos.
$\mathcal{M}$ can be diagonalized by a unitary matrix $V$ in the form of $ V^T\mathcal{M}V=\widehat{M}=\operatorname{diag}\{m_1,m_2,m_3,M_1,M_2,M_3\}  = \operatorname{diag}\{\widehat{M}_l, \widehat{M}_h\}$ for three generations of light and heavy neutrinos, where light neutrinos are the neutrinos we know in the SM and heavy neutrinos are the possible neutrinos yet to be found. The mass eigenstate neutrinos are given by  $ \psi^m_{L}=V^\dagger \psi_{L} $. 

In the mass eigenstate basis, assuming the momenta of all generation neutrinos are the same, the initially SM interaction produced active left-chiral neutrino states are
$ \psi_{Li}^h =\sum_{k}V_{ik}^*\psi^m_{Lk}$ and $ \psi_{Li}^h =\sum_{k}V_{ik}\psi^m_{Rk}$, where $k= 1 ... 6$. In the
above we have used $ (\psi^m_L)^c=((\psi^m)^c)_R $ and $ (\psi^m)^c=\psi^m $. 
$\psi_{Li}$ are the light neutrino for $i=1,2,3$ and the heavy neutrino for $i=4,5,6$, respectively.  We have
\begin{eqnarray}	\mathcal{L}=\frac{1}{2}\left( 
	\bar\psi^m (i\slashed{\partial}-\widehat{M}) \psi^m
	\right) 
	-\bar\psi^m\widetilde{J}^\mu\gamma_\mu\frac{1-\gamma_5}{2}\psi^m\;,
\end{eqnarray}
and the equation of motion is
\begin{equation}
	(i\slashed{\partial}-\widehat{M}) \psi^m-
	\widetilde{J}^\mu\gamma_\mu\frac{1-\gamma_5}{2}\psi^m
	+(\widetilde{J}^\mu)^*\gamma_\mu\frac{1+\gamma_5}{2}\psi^m=0
	\;.
\end{equation}
Here $\psi^m = \psi^m_L + (\psi^m_L)^c$ and $\widetilde{J}^\mu= V^\dagger J^\mu V$. So the Hamiltonian $H$ for the general case is
	\begin{eqnarray}
		\label{general Hamiltonian}
		H=(\mathbf{p}\cdot\boldsymbol{\Sigma})\gamma^5
		+ \beta \begin{pmatrix}
			\widehat{M}_l&0\\
			0&\widehat{M}_h
		\end{pmatrix}
		+ \widetilde{J}^\mu \gamma^0\gamma_\mu\frac{1-\gamma^5}{2}
		-\left(\widetilde{J}^\mu\right)^*\gamma^0\gamma_\mu\frac{1+\gamma^5}{2}\;.
	\end{eqnarray}
From the expression of Hamiltonian, we can easily see that the helicity is conserved when neutrinos pass through the matter.

From the above, in the frame where matter is at rest, one can easily recover the usual matter oscillation formalism with $J^\mu_L = (\rho, \mathbf{0})$ in the relativistic case $p\gg M>m\gg\rho$. Keeping the leading effect in this limit, one obtains,
\begin{equation}
	H_{\mathrm{eff}}=p+\frac{M^\dagger M}{2p}
	-h\cdot \rho\;.
\end{equation}
Then we can find that the matter effect would influence the contribution from mixing angle and mass square. 
Note that
for helicity $h=-1$, it is the usual  leading order neutrino oscillation in matter which can cause matter induced MSW resonant effect. But for $h=+1$ or $\rho<0$, the matter effects are different.

To explicit the chiral oscillation, let us consider the simple case of just one light active neutrino $\nu_L$ and one heavy sterile neutrino $N_R$ with seesaw mass matrix $\widehat{M} = \operatorname{diag}\{m,M\}$.
Then we can parameterize the mixing matrix as
\begin{equation}
	V=\begin{pmatrix}
		V_{a1}&V_{a2}\\
		V_{s1}&V_{s2}
	\end{pmatrix} 
	=\begin{pmatrix}
		\cos \theta & e^{i \eta}\sin \theta  \\
		-\sin \theta\;\;\; & e^{i \eta}\cos \theta 
	\end{pmatrix}\;.
\end{equation}
$\theta$ is the mixing angle between light and heavy neutrinos and $\eta$ is a Majorana phase which does not show up in the usual approximation for neutrino oscillation.

With $j^\mu_{R, RL} = 0$ for the case of having just the SM interactions in matter, we have
\begin{equation}
	\widetilde{J}^\mu\gamma_\mu= V^\dagger J^\mu V\gamma_\mu
	=\begin{pmatrix}
		\frac{\rho}{2}(1+\cos2\theta)&\frac{\rho}{2}e^{i\eta}\sin2\theta\\
		\frac{\rho}{2}e^{-i\eta}\sin2\theta&\frac{\rho}{2}(1-\cos2\theta)
	\end{pmatrix}\gamma_0\;.
\end{equation}
The Hamiltonian for this system becomes
	\begin{equation}
		\label{Dirac+Majorana Ham}
		H=\begin{pmatrix}
			\frac{\rho}{2}(1+\cos2\theta)-\mathbf{p}\cdot\boldsymbol{\sigma}
			&m&\frac{\rho}{2}e^{i\eta}\sin2\theta&0\\
			m&\mathbf{p}\cdot\boldsymbol{\sigma}-\frac{\rho}{2}(1+\cos2\theta)&0&-\frac{\rho}{2}e^{-i\eta}\sin2\theta\\
			\frac{\rho}{2}e^{-i\eta}\sin2\theta&0&
			\frac{\rho}{2}(1-\cos2\theta)-\mathbf{p}\cdot\boldsymbol{\sigma}&M\\
			0&-\frac{\rho}{2}e^{i\eta}\sin2\theta&M&\mathbf{p}\cdot\boldsymbol{\sigma}-\frac{\rho}{2}(1-\cos2\theta)
		\end{pmatrix}\;.
	\end{equation}

It is interesting to note that one can obtain analytic eigen-values of the above $H$ and therefore study the oscillation behaviors in details. The eigen-values are $E_{1h}=-\sqrt{A_1-A_2}$, $E_{2h}=\sqrt{A_1-A_2}$, $E_{3h}=-\sqrt{A_1+A_2}$ and $E_{4h}=\sqrt{A_1+A_2}$,
where
	\begin{equation}
		\label{eigen H}
		\begin{aligned}
			A_{1}=&\frac{m^2+M^2}{2}+\left(h\cdot p-\frac{\rho}{2}\right)^2+\frac{\rho^2}{4}\;,\\
			A_{2}=&\frac{\sqrt{
					\Big((m^2-M^2)\cos2\theta- 2 \rho (h\cdot p-\tfrac{\rho}{2}) \Big)^2
					+\Big(\left(m^2-M^2\right)^2
					+ \rho^2\left( m^2+M^2-2mM\cos2\eta\right)\Big)
					\sin^2 2\theta
			}}{2}\;.
		\end{aligned}
	\end{equation}
From the above one can obtain the evolution matrix $U(t)$. The elements in $U(t)$ are known analytic functions of $ m $, $ M $, $p$, $\rho$, $\theta$ and $\eta$. The important features we would like to mention are that there are resonant chiral oscillations for a given matter density and also the Majorana phase $\eta$ can affect chiral rotations. We discuss different cases in the following.

If one just considers two active Majorana neutrinos, such as $ \nu_e $ and $ \nu_\mu $ in Type-II seesaw model, we can get the results from the case of one active and one sterile neutrino directly with the replacement
\begin{equation}
	\label{replace}
	\nu_{L}\rightarrow\nu_{eL}\;,\;\;N_{R}^c\rightarrow\nu_{\mu L} \;,\;\;
	h\cdot p\rightarrow h\cdot p-\rho_Z \;,\;\; \rho\rightarrow \rho_W\;.
\end{equation}
We will give some examples in the following discussion.

Before discussing the general case, we discuss how the pure Dirac neutrino and type II seesaw neutrino oscillations are affected by matter effects.
Setting $\rho = 0$, $M_{L, R} =0$, and $M=m$, and combining $\nu_L + N_R$ to have a  Dirac state $\psi$, one recovers the free space Dirac neutrino oscillation case that we discussed in the previous parts. If $\rho$ is not zero, we have
$H = (\mathbf{p}\cdot\boldsymbol{\Sigma})\gamma^5+ m\beta+\rho (1-\gamma_5 )/2$
with eigen-values $E_1= \rho/ 2 + E_h$ and $E_2= \rho/2 - E_h$
where $ E_h = \sqrt{m^2 + (h\cdot p-\rho/2)^2} $,  and the corresponding eigen-states
\begin{equation}
\begin{aligned}
	\psi_{1}=&\frac{1}{\sqrt{2E_h}}
	\begin{pmatrix}
		\sqrt{E_h-(h\cdot p-\frac{\rho}{2})}\ u^h\\
		\sqrt{E_h+(h\cdot p-\frac{\rho}{2})}\ u^h
	\end{pmatrix}\;,\;\;
	\psi_{2}=\frac{1}{\sqrt{2E_h}}
	\begin{pmatrix}
		\sqrt{E_h+(h\cdot p-\frac{\rho}{2})}\ u^h\\
		-\sqrt{E_h-(h\cdot p-\frac{\rho}{2})}\ u^h
	\end{pmatrix}\;.
\end{aligned}
\end{equation}
$E_1$ and $E_2$ correspond to the positive and negative energy cases in free space propagation. Note that each of the original energy is split into two levels depending on the helicity of the states.

One can easily derive the oscillation probabilities for given helicity and chirality with
\begin{equation}
	\label{interaction Dirac amplitude}
	P(\nu_L^h\rightarrow {N}_R^h)
	=\frac{m^2}{E_h^2}\sin^2(E_h t)
	=1-P(\nu_L^{h}\rightarrow \nu_L^{h}).
\end{equation}
One can also obtain the similar results for a right-chiral initial state. From Eq.(\ref{interaction Dirac amplitude}) we can find that there's a resonance when $h\cdot p=\rho/2$.

Setting $M=0$, $M_D =0$, $j^{\mu}_{RL}=0$, and $ \theta=0 $, one obtains the pure active left-chiral Majorana neutrino case which can be realized in Type II seesaw model. In this case Hamiltonian $H$ of mass eigenstate $\psi=\nu_{L}+\nu_{L}^c$ is given by $H =(\mathbf{p}\cdot\boldsymbol{\Sigma})\gamma^5+ m\beta -\rho\gamma^5$. 
The positive eigen-energy is $E_h = \sqrt{m^2 + (h\cdot p-\rho)^2}$ and the negative one is $-E_h$.
Then the chiral oscillation amplitudes will be modified with $\rho/2$ replaced by $\rho$ in Eq.(\ref{interaction Dirac amplitude}),
and the expression of chiral oscillation $P(\nu^h_{L}\rightarrow \nu^h_{L})$ and $P(\nu^h_{L}\rightarrow (\nu^h_{L})^c)$ would be the same as Eq.(\ref{interaction Dirac amplitude}) 
with the replacement $N_R^h$ to $(\nu^h_{L})^c$.
The resonant enhanced chiral oscillation occurs at $h\cdot p = \rho$, which is different to Eq.(\ref{interaction Dirac amplitude}). 

In the usual case, $p\gg m\gg\rho$, the contribution of $\rho$ is very small so that Eq.(\ref{interaction Dirac amplitude}) degenerate to Eq.(\ref{free}) and Eq.(\ref{freee}). While in the dense matter, considering the case that $ p $ and $ \rho $ are comparable, the matter effect can make a large contribution to chiral oscillation. For example, considering the matter effect in the internal of neutron star, 
$\rho=-G_F N_n/\sqrt{2} =-(3.82\times10^{-14}\mathrm{eV})\cdot a/(\mathrm{g}/\mathrm{cm}^3)$
where the mass density $a$ can be as large as $10^{15} \mathrm{g/cm}^3$ \cite{Ozel:2016oaf}, leading to a not so small $\rho$. 
For $h=-1$ there's a resonant density when $\rho/2$ ({or} $\rho$) is equal to momentum for Dirac neutrino (or Type II seesaw Neutrino) leading to $E_h = m$. Then the chiral oscillation probability becomes the largest. While as for $h=+1$, there  is no such a resonant effect.

\section{The Majorana phase effects}
For neutrino chiral oscillation, the Majorana phases can also play a role unlike the usual neutrino oscillation. Let us study the simple case of general seesaw neutrino oscillation in Eq.(\ref{Dirac+Majorana Ham}). In vacuum, $\rho=0$, one obtains the general seesaw neutrino oscillation probabilities for one light and one heavy neutrinos in vacuum as 
	\begin{equation}
		\label{free prob}
		\begin{aligned}
			&P(\nu_{L}^h\rightarrow \nu_{L}^h)
			=\left(\cos^2\theta\cos(E_m t)
			+\sin^2\theta\cos(E_M t)\right)^2
			+p^2\left(\frac{\cos^2\theta}{E_m}\sin(E_m t)
			+\frac{\sin^2\theta}{E_M}\sin(E_M t)\right)^2\;,\\
			&P(\nu_{L}^h\rightarrow (N^h_{R})^c)
			=\frac{\sin^2 2\theta}{4}\left( \left(\cos(E_m t)-\cos(E_M t)\right)^2
			+p^2\left(\frac{\sin(E_m t)}{E_m}-\frac{\sin(E_M t)}{E_M}\right)^2\right)\;,\\
			&P(\nu_{L}^h\rightarrow (\nu_{L}^h)^c)
			=\frac{m^2}{(E_m)^2}\cos^4\theta\sin^2(E_mt)
			+\frac{mM}{2E_mE_M}\sin^2 2\theta\cos2\eta\sin(E_mt)\sin(E_Mt)
			+\frac{M^2}{(E_M)^2}\sin^4\theta\sin^2(E_Mt)\;,\\
			&P(\nu_{L}^h\rightarrow N_{R}^h)
			=\frac{\sin^2 2\theta}{4}\left(\frac{m^2}{(E_m)^2}\sin^2(E_m t)-\frac{2mM}{E_m E_M}\cos2\eta\sin(E_m t)\sin(E_Mt)
			+\frac{M^2}{(E_M)^2}\sin^2(E_Mt)\right)\;,
		\end{aligned}
	\end{equation}
where $E_m = \sqrt{m^2 + p^2}$ and $E_M = \sqrt{M^2+p^2}$.
The last two equations provide information about chiral rotation, where we can see that the Majorana phase $\eta$ appears. However, such effects may be challenging to observe, besides the need to detect low energy neutrinos, one also notes that the time averaged terms proportional to $\cos2\eta$ vanish, making detection even more difficult. 
This is also true with more generations in vacuum,
\begin{equation}
	P(\psi_{Li}^h\rightarrow\psi_{Lj}^h)
	=\sum_{k}\left|V_{ik}^*V_{jk} \right|^2\left(\frac{1}{2}+\frac{p^2}{2E_k^2}\right)\;,\;\;
	P(\psi_{Li}^h\rightarrow(\psi_{Lj}^h)^c)
	=\sum_{k}\left|V_{ik}^*V_{jk}^*\right|^2\frac{m_k^2}{2E_k^2}\;.
\end{equation}We would like to emphasize that the results in Eq.(\ref{free prob}) can also apply to the oscillation between two active Majorana neutrinos in Type-II seesaw model by replacing the parameters and $ \nu_{L}$, $ N_{R}^c $ to $\nu_{eL}$, $\nu_{\mu L} $, respectively.

One anticipates that the oscillation patterns in matter will be more complicated than those in free space. An interesting finding is that in this
case, the chiral oscillation pattern in matter dependence on $\eta$ does not vanish even time average is taken. We illustrate this with the limit $p\gg M>m\gg\rho$. In this case we have
	\begin{equation}
		\begin{aligned}
			&P(\nu^h_{L}\rightarrow (\nu^h_{L})^c)
			=\frac{(m^2-M^2)^2}{8A_2^2}\left(
			\frac{(m^2\cos^4\theta+M^2\sin^4\theta)}{p^2}
			+\frac{2\rho^2(2m^2\cos^4\theta+2M^2\sin^4\theta+mM\cos2\eta\sin^2 2\theta)}{(m^2-M^2)^2}
			\right .\\
			&\hspace{7cm}\left . 
			-\frac{\rho(4m^2\cos^6\theta-4M^2\sin^6\theta+mM\cos2\eta\cos2\theta\sin^2 2\theta)}{p(m^2-M^2)} 
			\right)\;,\\
			&P(\nu^h_{L}\rightarrow N^h_{R})
			=\frac{(m^2-M^2)^2\sin^2 2\theta}{32A_2^2}
			\left(\frac{m^2+M^2}{p^2}
			-\frac{2\rho(m^2+M^2-2mM\cos2\eta)\cos2\theta}{p(m^2-M^2)} \right .\\
			&\hspace{10.5cm}\left . +\frac{4\rho^2\left( m^2+M^2-2mM\cos2\eta\right)}{(m^2-M^2)^2}
			\right)\;.
		\end{aligned}
	\end{equation}
The probabilities for the other two oscillation modes are
\begin{equation}
	\label{usual}
	P(\nu^h_{L}\rightarrow \nu^h_{L})
	=\frac{1}{2}+\frac{\cos^2 2\theta_{\mathrm{eff}}}{2} - P(\nu^h_{L}\rightarrow (\nu^h_{L})^c)
	\;,\;\;
	P(\nu^h_{L}\rightarrow (N^h_{R})^c)
	=\frac{\sin^2 2\theta_{\mathrm{eff}}}{2} - P(\nu^h_{L}\rightarrow N^h_{R})\;,
\end{equation}
where $\cos2\theta_{\mathrm{eff}}
=((M^2-m^2)\cos2\theta+ 2 \rho (h\cdot p-\rho/2))/(2A_2)$. We do see the oscillation probabilities dependence on $ \eta $ even taking the time average. Note that this Majorana phase effect vanishes if any of the $m$, $M$ and $\theta$ is zero.

\section{Numerical analysis of neutrino chiral oscillation}
\begin{figure*}[htb]
	\centering
	\subfigure[$\rho>0$]{\includegraphics[width=3.0in]{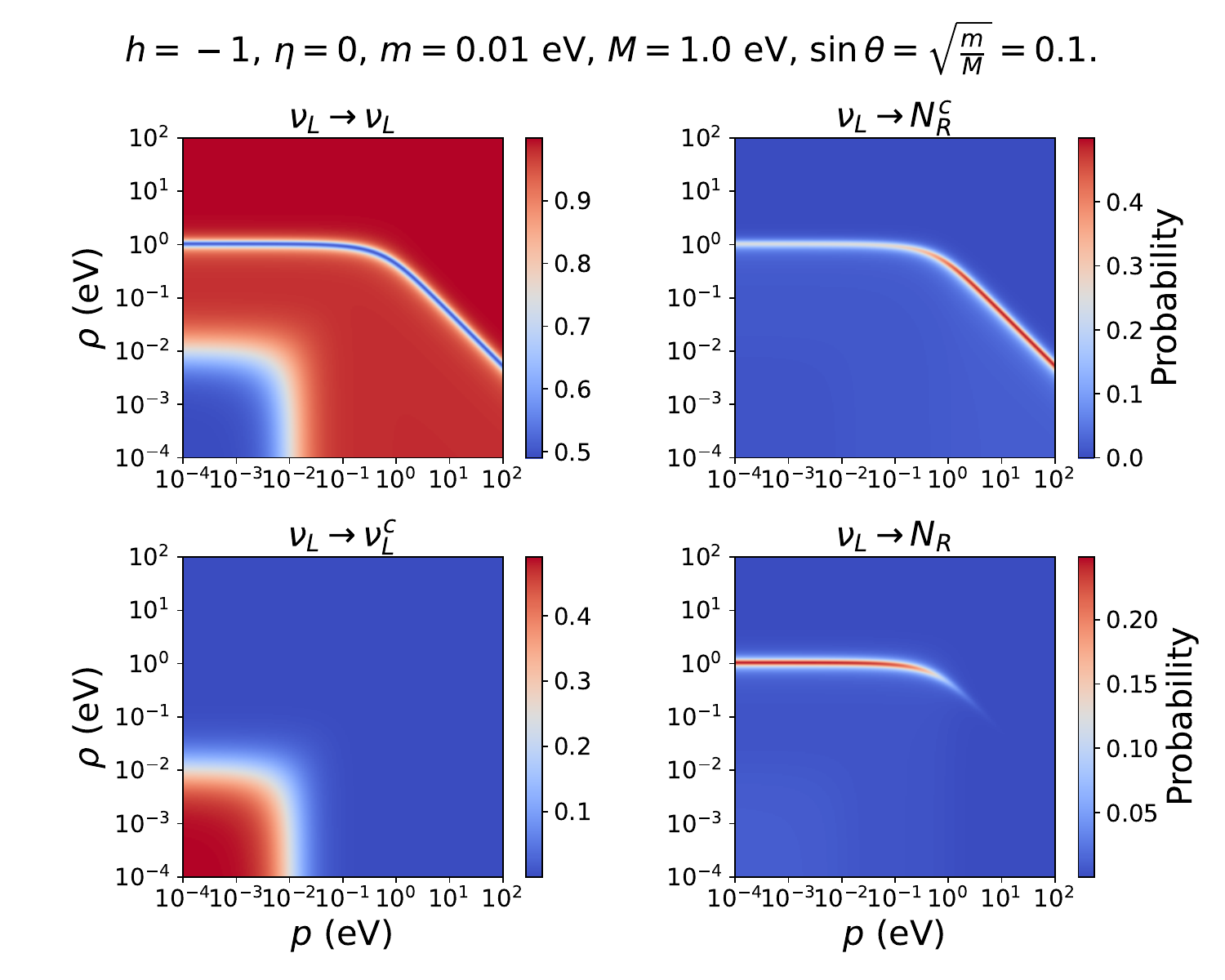}}\hspace*{0.1in}
	\subfigure[$\rho<0$]{\includegraphics[width=3.0in]{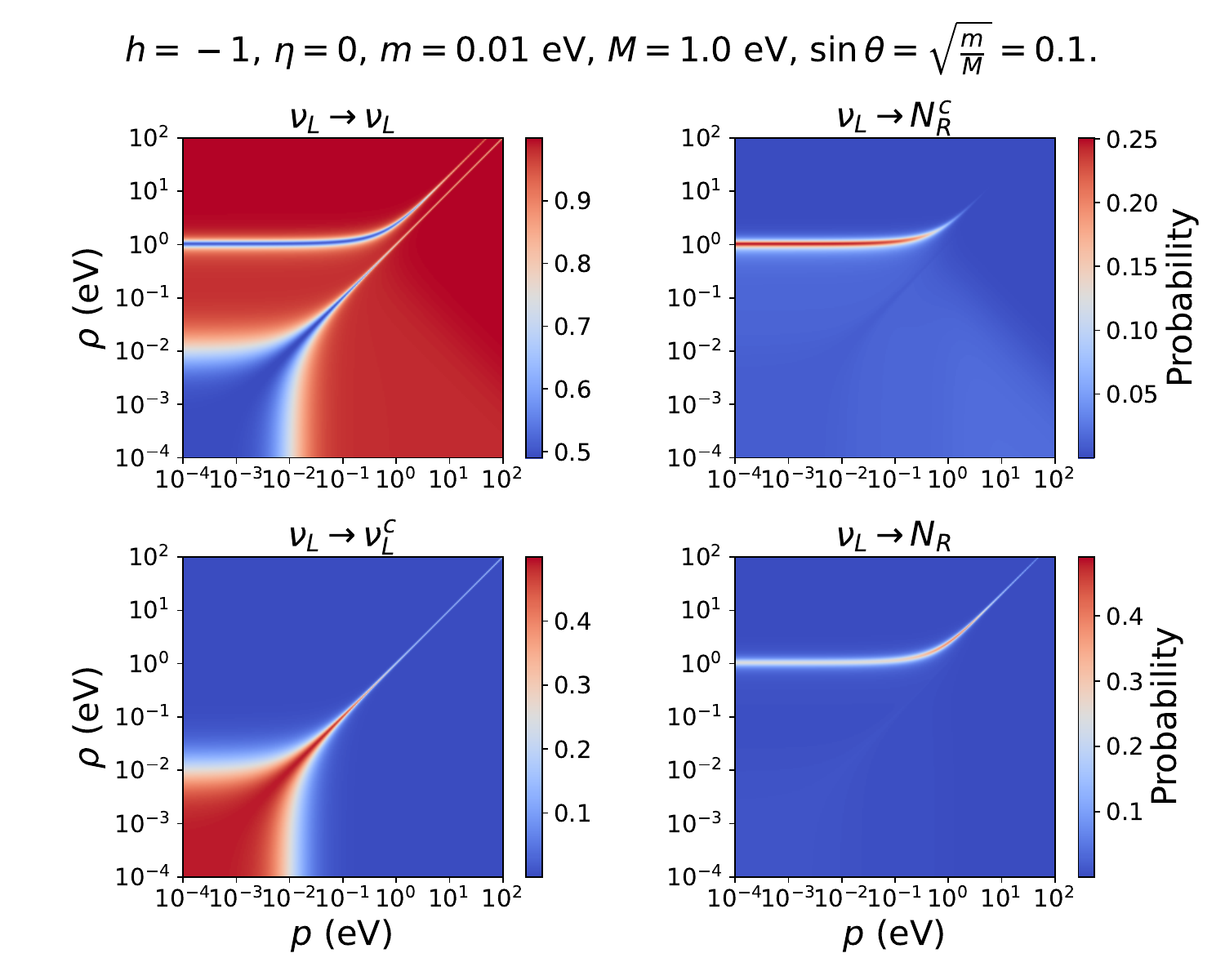}}
	\caption{\label{matter effect}The time average oscillation probabilities in matter for $\rho>0$ (a) and for $\rho <0$ (b) with $ \eta=0 $, and other input parameters are: $m=0.01\mathrm{eV}$, $M=1\mathrm{eV}$, and $\sin\theta=\sqrt{m/M}=0.1$.  
	}
\end{figure*}
\begin{figure*}[htbp] 
	\centering
	\includegraphics[width=6in]{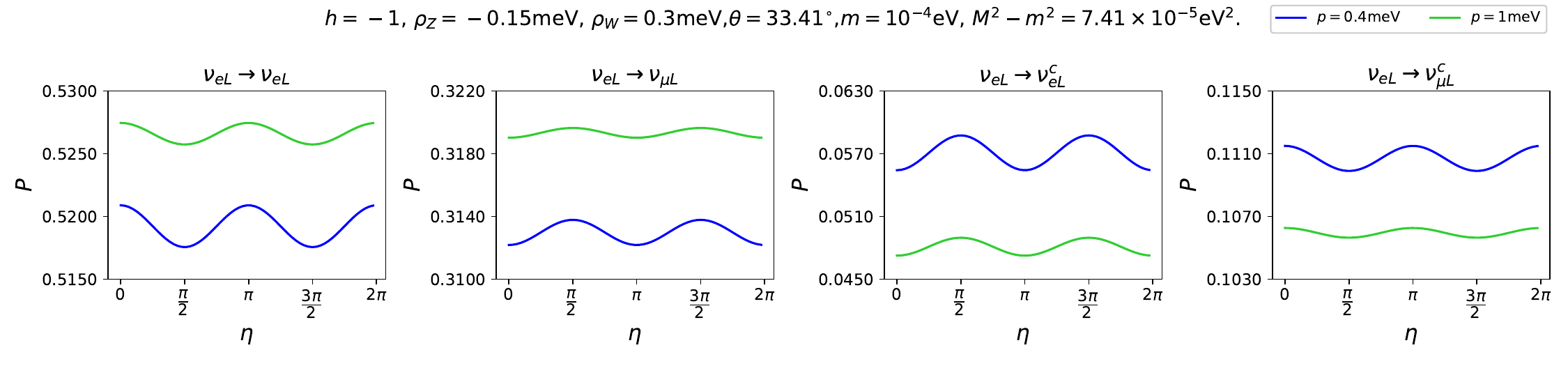}
	\caption{\label{Majorana12 fig}
		The time averaged Majorana phase $\eta$ effects on probabilities in two active Majorana relic neutrinos case for different neutrino momentum (0.4, 1.0) meV. In the vertical axis, $P$ is the time averaged probabilities in matter.
		The lighter neutrino mass is fixed to be $0.1$ meV.	The values of mixing angle $ \theta=\theta_{12} $ and mass square difference $M^2-m^2=\Delta m_{21}^2 $ are from \cite{ParticleDataGroup:2022pth}. For the matter densities, we use
		$\rho_W=0.3\mathrm{meV}$ and $\rho_Z=-0.15\mathrm{meV}$. Such a mater density could be found in neutron star cluster \cite{Chamel:2008ca}.}
\end{figure*}

We have carried a more detailed numerical analysis to see the matter and Majorana phase effects on the oscillation pattern in matter after taking the time average, which are shown in Fig. \ref{matter effect} and \ref{Majorana12 fig}.
In Fig. \ref{matter effect}, we take the type-I seesaw model as the inspiration to have Majorana neutrino masses. In that case the right-handed neutrinos are usually very heavy. However, some of the right-handed neutrino masses could be small as low as eV scale \cite{Giunti:2019aiy}, which allows for a relatively significant mixing angle $\theta$ between heavy and light neutrinos, resulting in a large effect on chiral oscillation. 
Therefore, we take $M=1\mathrm{eV}$, $m=0.01\mathrm{eV}$ in Eq.(\ref{Dirac+Majorana Ham}), to illustrate a relatively large mixing angle $\sin\theta=\sqrt{m/M}=0.1$.
Note that the parameter $m$ is the light neutrino mass which satisfies with the constraints from experiment \cite{Planck:2018vyg}.
In the $\nu_{L}\rightarrow \nu_L$ oscillation  picture, the blue region shows the smallest survival probability, which corresponds to the significant disappearance probability. And in the other three  pictures, the red region shows the considerable appearance probability. In the $\nu_{L}\rightarrow N_R^c$ oscillation picture of Fig. \ref{matter effect}(a), when energy is large enough, we can see the explicit matter effect in red region, which is consistent with the MSW effect for $\rho >0$. But for $\rho <0$ there is no MSW effect. 
Besides, in low energy region, we can see that the MSW effect would be suppressed, and the $\nu_{L}\rightarrow N_R$  oscillation would show the matter effect explicitly. As for low matter density and low energy region, the probability for $\nu^h_{L}\rightarrow (\nu^h_{L})^c $ oscillation becomes significant, which is consistent with the vacuum non-relativistic results.
It is interesting to note that there are two resonant regions in Fig. \ref{matter effect}(b). The resonant region in $\nu^h_{L}\rightarrow (\nu^h_{L})^c$ oscillation is consistent with the vacuum case before. While the resonant region in $\nu^h_{L}\rightarrow N_R$ oscillation is additional resonant effect. 

As the last illustrative example, we consider time average  oscillation probability of two active Majorana neutrinos, $ \nu_e $ and $ \nu_\mu $, to see how Majorana phase affect oscillation behavior, especially for CNB. The formulae can be obtained by replacement in Eq.(\ref{replace}) from the above and $M$ and $m$ are now the two tiny neutrino mass in the SM. The results for probability changes with $\eta$ are shown in Fig.\ref{Majorana12 fig} for  different neutrino momenta. The absolute variation ratio $R=|(P_{p=0.4\mathrm{ meV}}-P_{p=1.0\mathrm{ meV}})/P_{p=1.0\mathrm{ meV}}|$ can be as large as $\lbrace$1.2\%, 1.8\%, 20.0\%, 4.9\%$\rbrace$ from left to right pictures in Fig.\ref{Majorana12 fig}. 
One can also see that with the variation of  Majorana phase $\eta$ the change of $R$ can be as large as $\lbrace$0.3\%, 0.3\%, 2.7\%, 0.9\%$\rbrace$.
The choice of the momenta, 0.4 and 1.0 meV, are constrained by the need of having $p>\rho$ for neutrinos to pass through the media \cite{Pantaleone:1992xh} and have significant flux for CNB. Once CNB has been detected, a following careful study of its fluxes with different momentum will help to extract the information about Majorana phase. To this end we would like to mention that the PTOLEMY \cite{PTOLEMY:2018jst} to measure the CNB will be the first step toward this goal.
As we have pointed out in the introduction, since detection of neutrinos in experiments are detected by SM weak interaction,
left-chiral neutrino disappearance probability will be the way to see any chiral oscillation effects for Dirac neutrinos, and for Majorana neutrinos, because the charge conjugated left-chiral neutrinos are right-chiral ones, left-chiral to right-chiral neutrino appearance processes can also be measured.

The new effects obtained in the above have profound impacts on cosmic and astro neutrinos physics, and understanding neutrino properties, especially for relic neutrino. We will present more detailed phenomenological implications for neutrino chiral oscillation in cosmology and astrophysics elsewhere \cite{work}.

\section*{Acknowledgements}
This work was supported in part by Key Laboratory for Particle Physics, Astrophysics and Cosmology, Ministry of Education, and Shanghai Key Laboratory for Particle Physics and Cosmology (Grant No. 15DZ2272100), and in part by the NSFC (Grant Nos. 11735010, 11975149, 12090064 and 123B2079). XGH was supported in part by the MOST (Grant No. MOST 106-2112-M-002-003-MY3 ).


\begin{thebibliography}{99}
	%\cite{ParticleDataGroup:2022pth}
	\bibitem{ParticleDataGroup:2022pth}
	R.~L.~Workman \textit{et al.} [Particle Data Group],
	%``Review of Particle Physics,''
	PTEP \textbf{2022}, 083C01 (2022)
	doi:10.1093/ptep/ptac097
	%2573 citations counted in INSPIRE as of 29 Feb 2024
	
	%\cite{Bittencourt:2020xen}
	\bibitem{Bittencourt:2020xen}
	V.~A.~S.~V.~Bittencourt, A.~E.~Bernardini and M.~Blasone,
	%``Chiral oscillations in the non-relativistic regime,''
	Eur. Phys. J. C \textbf{81}, no.5, 411 (2021)
	doi:10.1140/epjc/s10052-021-09209-2
	[arXiv:2009.00084 [hep-ph]].
	%10 citations counted in INSPIRE as of 28 Feb 2024
	
	%\cite{Ge:2020aen}
	\bibitem{Ge:2020aen}
	S.~F.~Ge and P.~Pasquini,
	%``Parity violation and chiral oscillation of cosmological relic neutrinos,''
	Phys. Lett. B \textbf{811}, 135961 (2020)
	doi:10.1016/j.physletb.2020.135961
	[arXiv:2009.01684 [hep-ph]].
	%12 citations counted in INSPIRE as of 28 Feb 2024
	
	%\cite{Kimura:2021qlh}
	\bibitem{Kimura:2021qlh}
	K.~Kimura and A.~Takamura,
	%``New CP Phase and Exact Oscillation Probabilities of Dirac Neutrino derived from Relativistic Equation,''
	[arXiv:2101.03555 [hep-ph]].
	%6 citations counted in INSPIRE as of 28 Feb 2024
	
	%\cite{Bittencourt:2022hwn}
	\bibitem{Bittencourt:2022hwn}
	V.~A.~S.~V.~Bittencourt, A.~E.~Bernardini and M.~Blasone,
	%``Chiral oscillations,''
	EPL \textbf{139}, no.4, 44002 (2022)
	doi:10.1209/0295-5075/ac8446
	%4 citations counted in INSPIRE as of 28 Feb 2024
	
	%\cite{Blasone:2022fgj}
	\bibitem{Blasone:2022fgj}
	M.~Blasone, V.~A.~S.~V.~Bittencourt and A.~E.~Bernardini,
	%``Chiral oscillations in three-flavor neutrino mixing,''
	PoS \textbf{CORFU2021}, 065 (2022)
	doi:10.22323/1.406.0065
	%2 citations counted in INSPIRE as of 28 Feb 2024
	
	%\cite{Pantaleone:1992xh}
	\bibitem{Pantaleone:1992xh}
	J.~T.~Pantaleone,
	%``Dirac neutrinos in dense matter,''
	Phys. Rev. D \textbf{46}, 510-523 (1992)
	doi:10.1103/PhysRevD.46.510
	%187 citations counted in INSPIRE as of 28 Feb 2024
	
	%\cite{Wolfenstein:1977ue}
	\bibitem{Wolfenstein:1977ue}
	L.~Wolfenstein,
	%``Neutrino Oscillations in Matter,''
	Phys. Rev. D \textbf{17}, 2369-2374 (1978)
	doi:10.1103/PhysRevD.17.2369
	%5875 citations counted in INSPIRE as of 28 Feb 2024
	
	%\cite{Mikheyev:1985zog}
	\bibitem{Mikheyev:1985zog}
	S.~P.~Mikheyev and A.~Y.~Smirnov,
	%``Resonance Amplification of Oscillations in Matter and Spectroscopy of Solar Neutrinos,''
	Sov. J. Nucl. Phys. \textbf{42}, 913-917 (1985)
	%4129 citations counted in INSPIRE as of 28 Feb 2024
	
	%\cite{Ozel:2016oaf}
	\bibitem{Ozel:2016oaf}
	F.~\"Ozel and P.~Freire,
	%``Masses, Radii, and the Equation of State of Neutron Stars,''
	Ann. Rev. Astron. Astrophys. \textbf{54}, 401-440 (2016)
	doi:10.1146/annurev-astro-081915-023322
	[arXiv:1603.02698 [astro-ph.HE]].
	%1016 citations counted in INSPIRE as of 29 Feb 2024
	
	%\cite{Giunti:2019aiy}
	\bibitem{Giunti:2019aiy}
	C.~Giunti and T.~Lasserre,
	%``eV-scale Sterile Neutrinos,''
	Ann. Rev. Nucl. Part. Sci. \textbf{69}, 163-190 (2019)
	%doi:10.1146/annurev-nucl-101918-023755
	[arXiv:1901.08330 [hep-ph]].
	%171 citations counted in INSPIRE as of 18 May 2024
	
	%\cite{Planck:2018vyg}
	\bibitem{Planck:2018vyg}
	N.~Aghanim \textit{et al.} [Planck],
	%``Planck 2018 results. VI. Cosmological parameters,''
	Astron. Astrophys. \textbf{641}, A6 (2020)
	[erratum: Astron. Astrophys. \textbf{652}, C4 (2021)]
	doi:10.1051/0004-6361/201833910
	[arXiv:1807.06209 [astro-ph.CO]].
	%12958 citations counted in INSPIRE as of 29 Feb 2024
	
	%\cite{Chamel:2008ca}
	\bibitem{Chamel:2008ca}
	N.~Chamel and P.~Haensel,
	%``Physics of Neutron Star Crusts,''
	Living Rev. Rel. \textbf{11}, 10 (2008)
	doi:10.12942/lrr-2008-10
	[arXiv:0812.3955 [astro-ph]].
	%443 citations counted in INSPIRE as of 28 Feb 2024
	
	%\cite{PTOLEMY:2018jst}
	\bibitem{PTOLEMY:2018jst}
	E.~Baracchini \textit{et al.} [PTOLEMY],
	%``PTOLEMY: A Proposal for Thermal Relic Detection of Massive Neutrinos and Directional Detection of MeV Dark Matter,''
	[arXiv:1808.01892 [physics.ins-det]].
	%108 citations counted in INSPIRE as of 28 Feb 2024
	
	\bibitem{work}
	Worked in progress.
\end{thebibliography}
\end{document}